\documentclass[conference]{IEEEtran}

\makeatletter
\def\ps@headings{%
\def\@oddhead{\mbox{}\scriptsize\rightmark \hfil \thepage}%
\def\@evenhead{\scriptsize\thepage \hfil \leftmark\mbox{}}%
\def\@oddfoot{}%
\def\@evenfoot{}}
\makeatother
\usepackage{graphicx}
\usepackage{dcolumn}
\usepackage{bm}
\usepackage{epsfig}
\usepackage{psfrag}
\usepackage{subfigure}
\usepackage{amsmath, amssymb, latexsym,amsfonts,epsfig, graphicx, verbatim}
\usepackage[nospace]{cite}

\newtheorem{theorem}{Theorem}

\newcommand{\E}{\mathbb{E}}
\newcommand{\pr}{\mathbb{P}}

\newcommand{\T}{\mathcal{T}}

\newcommand{\R}{\mathcal{R}}

\newcommand{\Sch}{\mathcal{S}}
\newcommand{\tfec}{t_{\textrm{FEC}}}
\newcommand{\back}{\!\!\!}

\begin{document}

\title{Exploiting the Path Propagation Time Differences in Multipath Transmission with FEC}

\author{Maciej Kurant\\
LCA - School of Communications and Computer Science\\
EPFL, CH-1015 Lausanne, Switzerland
\\Email: maciej.kurant@epfl.ch}

\maketitle

\begin{abstract}
We consider a transmission of a delay-sensitive data stream from a single source to a single destination. The reliability of this transmission may suffer from bursty packet losses - the predominant type of failures in today's Internet. An effective and well studied solution to this problem is to protect the data by a Forward Error Correction (FEC) code and send the FEC packets over multiple paths.
\quad In this paper we show that the performance of such a multipath FEC scheme can often be further improved. Our key observation is that the propagation times on the available paths often significantly differ, typically by 10-100ms.
We propose to exploit these differences by appropriate packet scheduling that we call `Spread'. We evaluate our solution with a precise, analytical formulation and trace-driven simulations. Our studies show that Spread substantially outperforms the state-of-the-art solutions. It typically achieves two- to five-fold improvement (reduction) in the effective loss rate. Or conversely, keeping the same level of effective loss rate, Spread significantly decreases the observed delays and helps fighting the delay jitter.
\end{abstract}

\section{Introduction}
\begin{figure}[!t]
    \psfrag{P1}[c][]{\small $P_1, t_1$}
    \psfrag{P2}[c][]{\small $P_2, t_2$}
    \psfrag{P3}[c][]{\small $P_3, t_3$}
    \psfrag{s}[c][]{source $s$}
    \psfrag{rel}[r][]{\footnotesize relay node}
    \psfrag{d}[c][]{\small destination $d$}
    \psfrag{n1}[c][c][1][5]{\footnotesize $n_1$}
    \psfrag{n2}[c][c][1][30]{\footnotesize $n_2$}
    \psfrag{n3}[c][c][1][-25]{\footnotesize $n_3$}
    \psfrag{n}[c][]{\footnotesize $n$ FEC packets}
    \psfrag{k}[c][]{\footnotesize $k$ data packets}
    \psfrag{r}[c][]{\footnotesize $n\!-\!k$ redundancy packets}
    \psfrag{FEC}[c][]{\small FEC$(n,k)$}
    \begin{center}
\includegraphics[width=0.48\textwidth]{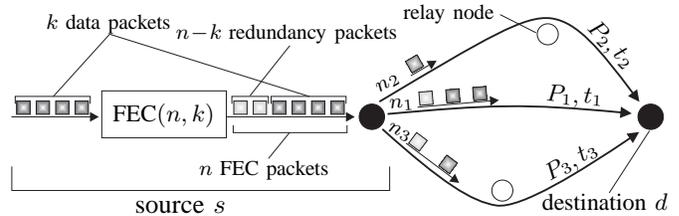}
\end{center}
\caption{Illustration of a multipath system with $R=3$ paths $P_1, P_2, P_3$ between source $s$ and destination $d$. $t_1,t_2,t_3$ are the corresponding path propagation times. $k$ data packets are complimented with $n-k$ redundancy packets, and the resulting $n$ FEC packets are split onto the three paths using the rates $n_1, n_2$ and $n_3$, respectively. }\label{fig:illustration}
\end{figure}

We consider a transmission of a delay-sensitive data stream from a single source to a single destination. How to improve the reliability of this transmission? Traditional ARQ (Automatic Repeat-reQuest) mechanisms often cannot be used, as they impose additional and usually unacceptable delays of at least one RTT (Round Trip Time). A more applicable technique is to introduce some type of redundancy, e.g., Forward Error Correction (FEC). Clearly, due to the delay constraints, a FEC block must be of limited length~\cite{Jiang02}. This, in turn, makes it inefficient against \emph{bursty packet losses}~\cite{Jiang02} - the predominant type of losses in today's Internet~\cite{Zhang01Constancy}. A good solution to this problem is to assign the FEC packets to \emph{multiple paths} spanning the source and the destination \cite{Maxemchuk75,Golubchik02,Nguyen03path,Yu05b,Vergetis05,Ribeiro05,Levy06,Li07SmartTunnel}.
An illustration of a multipath FEC system is presented in Fig.~\ref{fig:illustration}. Theoretically, the multiple paths could be constructed with the help of source routing, but this technique is not yet fully available in the Internet. A more practical alternative is the usage of overlay relay nodes that forward the traffic (as in Fig.~\ref{fig:illustration}). If the resulting paths are statistically independent, which is especially likely for multi-homed hosts, then the loss bursts get averaged out and FEC regains effectiveness. Similar performance benefits due to multipath were also observed in the context of Multiple Description Coding~\cite{Apostolopoulos01}.

When designing a system that splits a FEC block across multiple paths, we have to (1) select some paths out of all candidates, (2) assign the transmission rates to these paths, and (3) schedule the packets. The previous studies proposed techniques to solve (1-3) as a function of the statistical loss properties of the paths~\cite{Golubchik02,Nguyen03path,Li07SmartTunnel}.

However, there are
other important parameters affecting the performance of the multipath FEC system. In particular, in this paper we show that the propagation times on the available multiple paths often significantly differ. These differences, in turn, can be exploited to improve the system reliability. Below, we explain and motivate our approach on concrete examples and measurements.

\subsection{Propagation times on direct and indirect paths may differ significantly}

\begin{figure}[!t]
    \psfrag{pdf}[c][]{\small $pdf$}
    \psfrag{t2_t1}[c][]{\small $t_2-t_1$}
    \psfrag{t3_t1}[c][]{\small $\max(t_2,t_3)\!-\!t_1$}
    \psfrag{A}[r][]{\small $R=2$ paths}
    \psfrag{B}[r][]{\small $R=3$ paths}
    \begin{center}\includegraphics[width=0.5\textwidth]{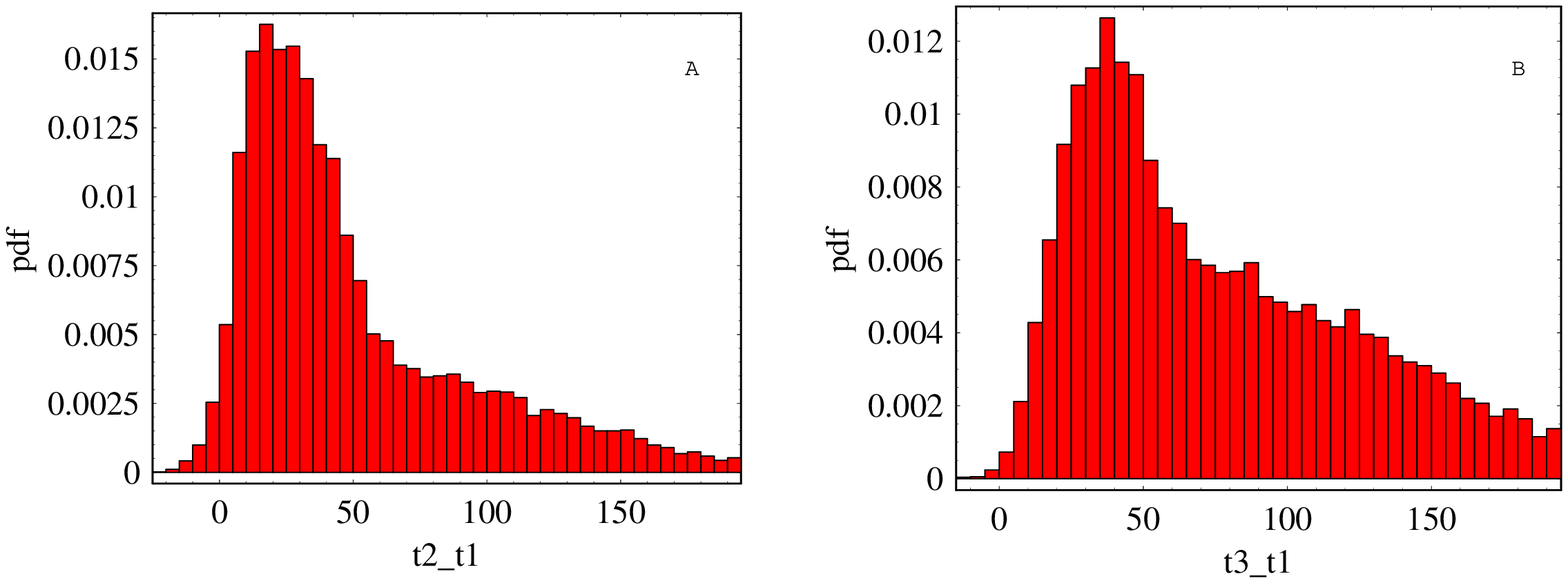}\end{center}
{\footnotesize
\begin{tabular}{|c|c|c|c|c|}
  \hline
   &&  C=1 & C=5 & C=50\\
   \hline
  $R=2$ & $t_2-t_1$ [ms]  & 43 & 37 & 18\\
  $R=3$ & $\back\max(t_2,t_3)\!-\!t_1$\back& N/A & 69 & 39\\
  \hline
\end{tabular}
}
\caption{The difference between propagation times on the direct path $P_1$ and the best indirect paths $P_2$ and $P_3$.
%
We present the results for $R=2$ (two paths: one direct and one indirect) and $R=3$ (the direct path and two indirect paths).
The histograms (top) show the distribution of propagation time differences for $C=5$ available candidate indirect paths. The table (bottom) shows the medians of these distributions for $C=1, 5$ and 50. The averages (not shown) are systematically higher than the medians.}\label{fig:indirect_path_delay}
\end{figure}

In Fig.~\ref{fig:indirect_path_delay} we study the path propagation time differences in the real-life Internet. The measurements were collected by running all-to-all traceroutes between 326 nodes in DIMES~\cite{WWW_netdimes}. These nodes are usually private hosts located at different sites around the world. (We obtained similar results for measurements on PlanetLab~\cite{WWW_PlanetLab}.)

For each source-destination pair we construct a set of $R$ paths. We always include the \emph{direct} path $P_1$ with propagation time $t_1$. Each of the remaining $R-1$ paths is \emph{indirect}, i.e., it uses some overlay relay node to forward the traffic. We choose uniformly at random a number $C$ of candidate relay nodes among the remaining 324 DIMES nodes. This results in $C$ candidate indirect paths.
From them we select the $R-1$ indirect paths following the intuitive selection procedure given in~\cite{Nguyen03path}. For $R\!=\!2$ paths we choose the indirect candidate path that is the most IP link disjoint with the direct path $P_1$. Clearly, this minimizes the loss correlation between $P_1$ and $P_2$. If there are more paths that achieve the minimal IP overlap, then the one with the smallest propagation time is kept. For $R>2$ we proceed similarly, except that now we consider the aggregated values of IP overlap and propagation time, i.e., summed over all $R(R-1)/2$ possible path pairs.




According to Fig.~\ref{fig:indirect_path_delay}, for $R=2$, the best indirect path $P_2$ has propagation times larger by typically $0\ldots 75ms$ than the direct path $P_1$ (see $t_2\!-\!t_1$ in top-left histogram). This difference gets larger for a smaller number of candidates $C$ (table at the bottom).

Moreover, the path propagation time differences grow significantly with the number of paths $R$ used in the system. As shown in Fig.~\ref{fig:indirect_path_delay}, already for $R\!=\!3$ the medians of the distributions are roughly doubled compared to $R\!=\!2$, and typically $P_1$ is faster than the slower of the two indirect paths by $\max(t_2,t_3)\!-\!t_1\simeq 0\dots 150ms$.

%

We conclude that in the real-life Internet the propagation time differences on multiple paths between a source-destination pair are significant, typically reaching several tens of milliseconds.

\subsection{The differences in propagation times can be exploited by a multipath FEC system}

\begin{figure}[!t]
    \psfrag{P1}[c][]{\small $P_1$}
    \psfrag{P2}[c][]{\small $P_2$}
    \psfrag{NOFEC}[c][]{}
    \psfrag{A}[l][]{\small A) \ delays}
    \psfrag{B}[l][]{\small B) \ no FEC}
    \psfrag{C}[l][]{\small C) \ $\Sch^{imm}_{(6,0)}$}
    \psfrag{D}[l][]{\small D) \ $\Sch^{imm}_{(3,3)}$ }
    \psfrag{E}[l][]{\small E) \ $\Sch^{spr}_{(3,3)}$}
    \psfrag{F}[l][]{\small F) \ $\Sch^{spr}_{(4,2)}$}
    \psfrag{LB}[l][]{$\pi^*_B=1.00\%$}
    \psfrag{LC}[l][]{$\pi^*_B=0.553\%$}
    \psfrag{LD}[l][]{$\pi^*_B=0.148\%$}
    \psfrag{LD1}[l][c][0.8]{(state of the art)}
    \psfrag{LE}[l][]{$\pi^*_B=0.113\%$}
    \psfrag{LF}[l][]{$\pi^*_B=0.016\%$}
    \psfrag{LF1}[l][c][0.8]{(proposed solution)}
    \psfrag{T3}[c][c][0.8]{$\Delta t=50ms$}
    \psfrag{START}[c][]{\small Start}
\includegraphics[width=0.38\textwidth]{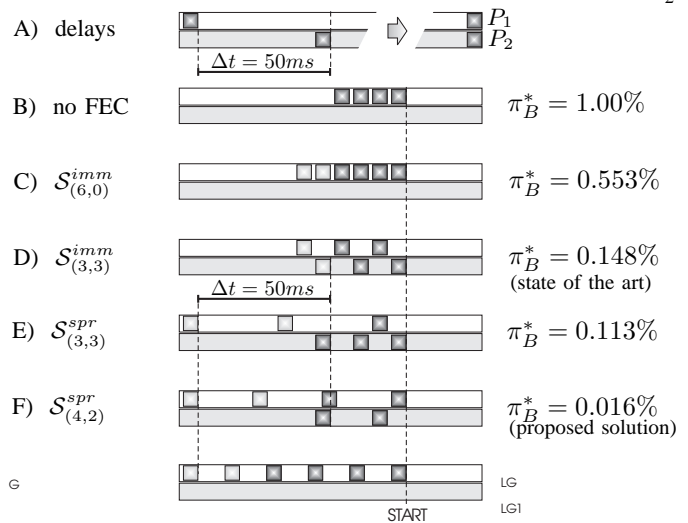}
\caption{Illustration of various packet schedules and their performance measured in the effective loss rate $\pi_B^*$.  We use two independent paths $P_1$ and $P_2$ with identical failure distributions. The data packets are generated at the source every $T=5ms$ and coded with FEC(6,4). \quad (A)~The path propagation time $t_2$ on path $P_2$ is $\Delta t\!=\!t_2\!-\!t_1\!=\!50ms$ larger than the path propagation time $t_1$ on path $P_1$.
\quad (B)~No FEC, single path, the packets are sent at times $0,5,10,15ms$. \quad (C)~FEC on $P_1$ only, packets are sent as soon as they are generated, i.e., we use the `Immediate' schedule $\Sch^{imm}$. \quad (D)~Packets alternate between $P_1$ and $P_2$ with equal rates rates $n_1\!=\!n_2\!=\!3$, as in~\cite{Nguyen03path,Li07SmartTunnel}. The total
FEC block
delay resulting from this scheme serves as a maximal FEC block delay in the following scenarios.
\quad (E)~Packets alternate between $P_1$ and $P_2$ with equal rates, but the three packets sent on $P_1$ are maximally spread. \quad (F)~Packets are split between $P_1$ and $P_2$ with optimal rates $n_1=4, n_2=2$, maximally spread.}\label{fig:schedulings}
\end{figure}

We propose to exploit these path propagation time differences when designing a multipath FEC system. Our solution is easy to implement and can bring significant performance gains. Let us take a concrete example described in Fig.~\ref{fig:schedulings}. Assume that there exist two paths between the source and the destination, the direct path $P_1$, and an indirect path $P_2$ created by employing another peer that works as a relay. Let $t_1=100ms$ and $t_2=150ms$ be the propagation delays on $P_1$ and $P_2$, respectively. So the path propagation time difference is $\Delta t=50ms$ (Fig.~\ref{fig:schedulings}A).
Let the two paths be lossy with the same loss rate $1\%$ and the same average loss burst length of $10ms$, but independent. The data packets are generated at the source every $T=5ms$. If no form of packet protection is used, then the data packet loss rate observed at the destination, or the \emph{effective loss rate}, is $\pi^*_B\!=\!1\%$ (B). Assume now that we use FEC(6,4) to protect the packets. If we send all packets on $P_1$ with inter-packet times $T$, then the effective loss rate after FEC decoding is $\pi^*_B\!=\!0.553\%$ (C). Following \cite{Nguyen03path,Li07SmartTunnel}, we now split the packets equally between $P_1$ and $P_2$, which decreases $\pi^*_B$ to $0.148\%$ (D). This solution represents the state of the art. Note that now the last FEC packet on path $P_2$ reaches the destination $\tfec=t_2\!+\!4\cdot T\!=\!170$ milliseconds after the generation of the first FEC packet at source. In other words, in this case the application using multipath FEC must accept the (maximal) delay equal to $\tfec$.
However, we can achieve far better results still respecting this delay constraint.
For instance, we can appropriately increase the packet-spacing on $P_1$ and achieve $\pi^*_B\!=\!0.113\%$ (E). Finally, we get even more significant improvement by sending 4 packets on $P_1$ and 2 packets $P_2$, i.e., by applying \emph{unequal} rates on the paths (F). This results in $\pi^*_B\!=\!0.016\%$, which is almost one order of magnitude smaller than (D).

In other words, we exploit the differences in path propagation times by spreading the packets in time, such that the maximal allowed delay is respected. The gain over the state of the art measured in the effective loss rate $\pi^*_B$ may be very significant (here $0.016\%$ vs $0.148\%$, i.e., almost ten-fold). Moreover, some results may seem counterintuitive. For instance, it may be better to use only one path than to use two (un-spaced) paths. It also turns out that even if the loss distributions on the paths are the same, the optimal rates assigned to these paths are not necessarily equal.


\subsection{Organization of this paper}
The remainder of this paper is organized as follows. In Section~\ref{sec:Notation and model} we fully specify our model, which allows us to precisely state the problem we are solving. Next, in Section~\ref{sec:Derivation of pi_B^*} we derive exact analytical expressions for the effective loss rate $\pi_B^*$ under multipath FEC and an arbitrary schedule. In Section~\ref{sec:Packet_scheduling} we describe the `Immediate' schedule representing the state of the art, and propose a `Spread' schedule that exploits the differences in path propagation times. In Section~\ref{sec:Performance evaluation} we evaluate our solution analytically, by simulations and by trace-driven simulations fed with real-life Internet traces. In Section~\ref{sec:Related work} we discuss the related work. Finally we conclude the paper and propose future directions. The details of some calculations are put in Appendix.

\section{Model and problem statement}\label{sec:Notation and model}
The packets, called \emph{data} packets, are generated at source~$s$, with constant inter-arrival time $T$. There exist $R$ paths between sender $s$ and destination $d$, with the propagation delays $t_1,\ldots,t_R$, respectively.

\subsection{Path losses}
The paths are assumed to be independent. We model bursty losses on each path by the popular two-state Gilbert model. Its basic version is a Discrete Time Markov Chain (DTMC), and captures the loss correlations due to queuing on bottleneck links, when the path is sampled at some constant rate (e.g., $1/T$). However, as we vary the sampling rates, DTMC is not sufficient. Indeed, on the same path we experience much higher loss burstiness under the packet interval of $T=5ms$ than of $T=100ms$~\cite{Jiang02}. For this reason
we use the continuous-time version of the Gilbert model~\cite{Bolot99,Golubchik02} that
naturally accommodates different sampling rates. It is a two-state stationary Continuous Time Markov Chain (CTMC) $\{X_r(t)\}$. The state $X_r(t)$ at time $t$ assumes one of the two values: $G$ (`good') or $B$ (`bad'). If a packet is sent at time $t$ and $X_r(t)=G$ then the packet is transmitted; if $X_r(t)=B$ then the packet is lost.

We denote by $\pi^{(r)}_G$ and $\pi^{(r)}_B$ the stationary probabilities that the $r$th path is good or bad, respectively. Similarly, let $\mu^{(r)}_G$ and $\mu^{(r)}_B$ be the transition rates from $G$ to $B$ and from $B$ to $G$, respectively. In this paper we use two meaningful, system-dependent parameters to specify the CTMC packet loss model:
\begin{itemize}
  \item the average loss rate $\pi^{(r)}_B$, and
  \item the average loss burst length (in seconds) $1/\mu^{(r)}_B$.
\end{itemize}
All other parameters can be easily derived from these two, because
\begin{equation}\label{eq:stationary_probabilities}
  \pi^{(r)}_G =\frac{\mu^{(r)}_B}{\mu^{(r)}_G+\mu^{(r)}_B}\quad \textrm{ and }\quad \pi^{(r)}_B =\frac{\mu^{(r)}_G}{\mu^{(r)}_G+\mu^{(r)}_B}.
\end{equation}

\begin{table}
  \centering
{\footnotesize
\begin{tabular}{|p{0.15\columnwidth}|p{0.75\columnwidth}|}
  \hline
  $\pr,\ \E$ & probability, \ expected value \\
  $s$ & source node \\
  $d$ & destination node \\
  $T$ & (constant) interval between two consecutive data packets at source $s$ \\
  $R$ & number of independent paths between source $s$ and destination $d$ \\
  $P_r$ & $r$th path\\
  $t_r$ & propagation delay on $P_r$ \\
  $\pi_B^{(r)}, 1\!/\!\mu_B^{(r)}$ & the average loss rate and loss burst length on path $P_r$\\
  $n,k,(n\!-\!k)$ & the number of FEC, data, and redundancy packets in a FEC block, respectively\\
  $n_r$ & number of FEC packets assigned to $P_r$ (rate of path $P_r$)\\
  $k_r$ & number of data packets assigned to path $P_r$ \\
  $T_r$ & (constant) spacing of the $n_r$ packets on path $P_r$\\
  $F,D$  &number of lost FEC and data packets before FEC recovery\\
  $\pi_B^*$ & effective loss rate, i.e., the expected fraction of lost data packets at the destination after the FEC recovery\\
  $\tfec$ & FEC block transmission time, i.e., the time between the generation of the first FEC packet at source $s$ and the scheduled delivery of the latest FEC packet at destination~$d$\\
  $\Sch\!=\!(\T,\!\R)$ & packet scheduling:  The $i$th packet in a FEC block is sent at time $\T(i)$ over path $\R(i)$\\
  \hline
\end{tabular}
}
\caption{Basic notation used in this paper.}\label{Tab:notation}
\end{table}

\subsection{Multipath FEC}

We use a FEC$(n,k)$ scheme to protect the data packets against losses (see Fig.~\ref{fig:illustration}). This means that $k$ data packets (not necessarily consecutive) are encoded as one FEC block of $n$ packets, called \emph{FEC} packets. In particular, as in~\cite{Frossard01,Golubchik02,Yu05,Yu05b,Li07SmartTunnel}, we consider a \emph{systematic}\footnote{The non-systematic FEC is easier to handle, but also less efficient. We show its analysis in Appendix.} FEC, i.e., a scheme where the first $k$ packets are the $k$ data packets (unchanged). The remaining $n-k$ packets, called \emph{redundancy} packets, carry the redundancy information. The destination uses the redundancy packets to recover some of the lost data packets as follows. Let $F$ be the number of lost FEC packets and let $D$ be the number of lost data packets of a FEC block, both before the FEC recovery (note that $D$ contributes to $F$).
If $F\leq n-k$ then all the $n$ FEC packets and hence all the $k$ data packets are recovered. In contrast, if $F> n-k$, then no FEC recovery is possible and $D$ data packets are lost.

\subsection{Packet scheduling}
\begin{figure}[!t]
    \psfrag{t1}[r][b][1][45]{\footnotesize $\T(1)\!=\!0$}
    \psfrag{t2}[r][b][1][45]{\footnotesize $\T(2)\!=\!T$}
    \psfrag{t3}[r][b][1][45]{\footnotesize $\T(3)$}
    \psfrag{t4}[r][b][1][45]{\footnotesize $\T(4)$}
    \psfrag{t5}[r][b][1][45]{\footnotesize $\T(5)$}
    \psfrag{t6}[r][b][1][45]{\footnotesize $\T(6)$}
    \psfrag{T}[c][c]{\small $T$}
    \psfrag{S}[l][c]{\small \parbox[c]{3cm}{Data packets generated at the source}}
    \psfrag{P1}[c][c]{\small $P_1$}
    \psfrag{P2}[c][c]{\small $P_2$}
    \psfrag{Q}[l][c]{\small \parbox[c]{4cm}{Schedule $\Sch=(\T,\R),$\\$\R=(2,1,1,2,1,1)$}}
    \psfrag{time}[c][c]{\small Time}
    \psfrag{FEC}[l][c]{\small \parbox[c]{4cm}{FEC(6,4)\\$n_1\!=\!4, k_1\!=\!2$\\$n_2\!=\!2,k_2\!=\!2$}}
    \begin{center}\includegraphics[width=0.5\textwidth]{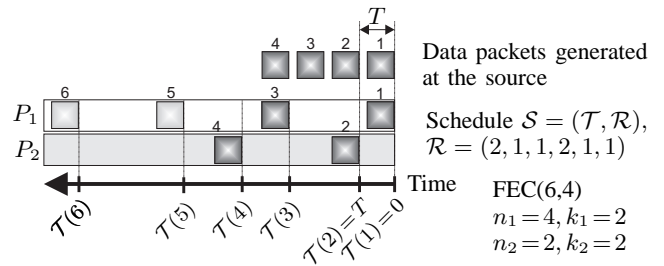}\end{center}
\caption{An illustration of a schedule $\Sch=(\T,\R)$ on $R\!=\!2$ paths with FEC(6,4). Four data packets numbered 1-4 are generated at the source at equal intervals~$T$; the first one specifies time $t=0$. The $n-k=2$ redundancy packets are numbered 5 and 6. According to the schedule $\Sch=(\T,\R)$, the $i$th FEC packet is sent at time $\T(i)\geq 0$ over path $\R(i)$.}\label{fig:schedule_illustration}
\end{figure}

Finally, the packets are sent according to some \emph{schedule} that defines \emph{when} and \emph{on which path} each FEC packet is sent. More precisely, we denote by $\Sch = (\T,\R)$ the schedule of packets in a FEC block, where $\T$ and $\R$ are vectors of length~$n$. The $i$th FEC packet is sent at time $\T(i)$ over path $\R(i)$, as shown in Fig.~\ref{fig:schedule_illustration}. The time is counted from the generation (at the source) of the first data packet of the FEC block. Denote by $\tfec$ the \emph{FEC block transmission time}, i.e., the time between the generation of the first FEC packet at source~$s$ and the scheduled delivery of the latest FEC packet at destination~$d$.
Given a schedule $\Sch$, $\tfec$ can be easily computed as
\begin{equation}\label{eq:t_fec}
    \tfec= \max_{1\leq i\leq n} \Big(\T(i)+t_{\R(i)}\Big).
\end{equation}
For a given schedule, $\tfec$ can be interpreted as the total delay imposed by the multipath FEC system on the delay-sensitive application using it. Indeed, if the first packet of a FEC block is lost and needs to be reconstructed by FEC, then we have to wait up to $\tfec$ until the destination is reached by the other FEC packets necessary for the reconstruction of the lost packet. In practice, however,
a constraint is likely to come from the delay-constrained application itself, as the maximal acceptable delay $\tfec$. In this case our goal is to design a good schedule respecting this constraint, which is the approach used in this paper.


The schedule also implicitly defines the \emph{rate} $n_r$ of path $P_r$, i.e., the number of FEC packets sent on $P_r$. Similarly, let $k_r$ be the number of data packets among the $n_r$ packets sent on $P_r$. Clearly, $\sum_{r}n_r = n$ and $\sum_{r}k_r = k$.

\subsection{Effective loss rate $\pi_B^*$ and problem statement}
Our ultimate goal is to send a stream of data packets over (possibly multiple) lossy channels in a way that minimizes the losses observed at the destination, given a maximal value for $\tfec$. Therefore, we adopt a natural performance metric called \emph{effective loss rate} $\pi_B^*$. It is defined as the expected fraction of lost data packets observed at the destination $d$ after an attempt of FEC decoding. Now the problem can be stated as follows:

\emph{Given the path loss properties ($\pi^{(r)}_B$, $1/\mu^{(r)}_B$ and $t_r$ for every path $P_r$), the FEC parameters ($n$ and $k$) and maximal FEC block transmission time $\tfec$, find the schedule $\Sch$ that minimizes the effective loss rate $\pi_B^*$.}

We approach this problem in two steps. First, in Section~\ref{sec:Derivation of pi_B^*} we derive an exact analytical formula for the effective loss rate $\pi_B^*$ for a given schedule $\Sch$. Second, in Section~\ref{sec:Packet_scheduling} we introduce a schedule that exploits the differences in path propagation times and outperforms the schedules proposed to date.

%
%

\section{Exact analytical derivation of the effective loss rate $\pi_B^*$} \label{sec:Derivation of pi_B^*}
In order to design a good schedule we must be able to evaluate it. In this section we derive the exact analytical expression for the effective loss rate $\pi_B^*$ for a given schedule~$\Sch$. We consider two cases. First, we derive $\pi_B^*$ for an arbitrary schedule~$\Sch$. The resulting formula is simple but computationally expensive and untractable for larger sizes $n$ of the FEC block. Next, we derive $\pi_B^*$ assuming that on each path separately the packets are evenly spaced. This constraint is compatible with the schedule we propose later and results in a computationally lighter formula for $\pi_B^*$.


\subsection{The effective loss rate $\pi_B^*$ for an arbitrary schedule}\label{subsec: effective loss rate for an arbitrary schedule}
Let $c$ be a $n$-tuple representing a particular failure configuration; $c_i,\, 1\!\leq\!i\!\leq\!n,$ takes the value $G$ (resp., $B$) if $i$th FEC packet is transmitted (resp., lost). By considering all possible failure configurations $c$ we can compute the effective loss rate $\pi_B^*$ for a given schedule $\Sch$ as follows:
\begin{equation}\label{eq:pi_B^* for arbitrary schedule, very short}
    \pi_B^*\ =\ \frac{1}{k} \sum_{\textrm{all }c} D(c)\cdot \pr(c),
\end{equation}
where $D(c)$ is the number of lost data packets (after the FEC recovery) for a given failure configuration $c$. For a systematic FEC$(n,k)$ we have
$$D(c) = \left\{ \begin{array}{ll}
    0 & \textrm{ if $\sum_{i=1}^n 1_{\{c_i=B\}}\leq n-k$}\\
    \sum_{i=1}^k 1_{\{c_i=B\}}& \textrm{ otherwise.}
    \end{array} \right. $$
In order to compute the probability $\pr(c)$ of a failure configuration $c$, we consider the $R$ paths separately, as follows. Denote by $\T^{(r)}$ the vector of length $n_r$ with departure times of packets scheduled by $\Sch$ on path $P_r$. Similarly, let $c^{(r)}$ be an $n_r$-element vector with the failure configuration on path $P_r$ defined by $c$. As the $R$ paths are independent, the probability $\pr(c)$ is
\begin{equation}\label{eq:pr(c) for arbitrary schedule, very short}
  \pr(c) = \prod_{r=1}^R \pr(c^{(r)}),
\end{equation}
where $\pr(c^{(r)})$ is the probability of a failure configuration $c^{(r)}$ on path $P_r$. The derivation of $\pr(c^{(r)})$ for the Continuous Time Markov Chain loss model is straightforward. Indeed, denote by $p^{(r)}_{i,j}(\tau)$ the probability of transition from state $i$ to state $j$ on path $P_r$ in time $\tau$, i.e.,
$$p^{(r)}_{i,j}(\tau) = \pr[X_r(\tau)=j | X_r(0)=i].$$
From classical Markov Chain analysis we have:
\begin{equation}\label{eq:p^r_i,j(tau)}
    \begin{array}{rcl}
p^{(r)}_{G,G}(\tau) &=& \pi^{(r)}_G+\pi^{(r)}_B\cdot \alpha \\
p^{(r)}_{G,B}(\tau) &=& \pi^{(r)}_B-\pi^{(r)}_B\cdot \alpha \\
p^{(r)}_{B,G}(\tau) &=& \pi^{(r)}_G-\pi^{(r)}_G\cdot \alpha \\
p^{(r)}_{B,B}(\tau) &=& \pi^{(r)}_B+\pi^{(r)}_G\cdot \alpha
\end{array}
\end{equation}
where $\alpha = \exp\big(-(\mu^{(r)}_G+\mu^{(r)}_B)\tau\big)$. Now $\pr(c^{(r)})$ can be easily computed. For example, for $c^{(r)}\!=\!GBB$ we have
$$\pr(c^{(r)}\!=\!GBB) = \pi^{(r)}_G\cdot p^{(r)}_{G,B}(\tau_1)\cdot p^{(r)}_{B,B}(\tau_2),$$
where $\tau_i$ is the time interval between the $i$th and $(i\!+\!1)$th FEC packet scheduled by $\Sch$ on path $P_r$, i.e, $\tau_i=\T^{(r)}_{i+1}-\T^{(r)}_i$. More generally,
%
%
%
\begin{equation}\label{eq:pr(c^{(r)}) for arbitrary schedule, long}
 \pr(c^{(r)})\ =\ \pi^{(r)}_{c^{(r)}_1} \prod_{i=1}^{n_r-1} p^{(r)}_{c^{(r)}_i,c^{(r)}_{i+1}}(\T^{(r)}_{i+1}-\T^{(r)}_i).
\end{equation}
Finally, we plug (\ref{eq:pr(c^{(r)}) for arbitrary schedule, long}) and (\ref{eq:pr(c) for arbitrary schedule, very short}) to (\ref{eq:pi_B^* for arbitrary schedule, very short}), to obtain
\begin{equation}\label{eq:pi_B^* brute force}
    \pi_B^*=\frac{1}{k} \sum_{\textrm{all }c} D(c) \prod_{r=1}^R \pi^{(r)}_{c^{(r)}_1} \prod_{i=1}^{n_r-1} p^{(r)}_{c^{(r)}_i,c^{(r)}_{i+1}}(\T^{(r)}_{i+1}-\T^{(r)}_i).
\end{equation}

\subsection{The effective loss rate $\pi_B^*$ for even spacing on paths}\label{subsec:Performance of FEC on independent paths}

\begin{figure}[!t]
    \psfrag{timeRatio}[c][c]{time\_ratio}
    \psfrag{n}[c][c]{$n$ - size of a FEC block}
    \begin{center}\includegraphics[width=0.5\textwidth]{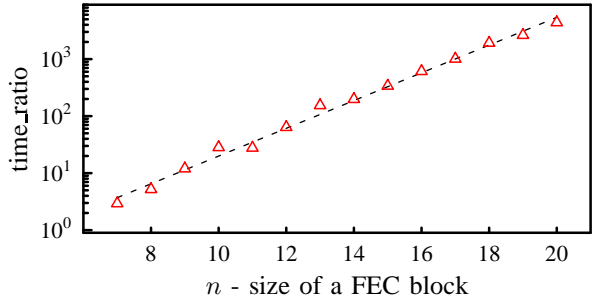}\end{center}
\caption{The time complexity of the effective loss rate $\pi_B^*$ under an arbitrary schedule (\ref{subsec: effective loss rate for an arbitrary schedule}) vs. the even-spaced schedule (\ref{subsec:Performance of FEC on independent paths}): time\_ratio is the runtime of Eq.~(\ref{eq:pi_B^* brute force}) divided by the runtime of Eq.~(\ref{eq:BIG pi_B*}). Here we use FEC$(n,0.7n)$ on two identical paths.}
\label{fig:timeComplexityOfExactEffectiveLossRate}
\end{figure}

Equation~(\ref{eq:pi_B^* brute force}) allows us to compute the effective loss rate $\pi_B^*$ for any schedule $\Sch$. However, evaluating~(\ref{eq:pi_B^* brute force}) is computationally expensive because the main sum is over all $2^n$ failure configurations. Thus it can be applied to relatively small $n$ only.
Fortunately, we can significantly reduce the computation complexity by assuming that on each path $P_r$ separately the packets are \emph{evenly spaced}, i.e., for all $1\!\leq\!i\!\leq\!n_r\!\!-\!\!1$ the intervals $\T^{(r)}_{i+1}-\T^{(r)}_i$ are the same and equal to a constant that we denote by $T_r$. Indeed, this constraint leads us to a formulation of $\pi_B^*$ (below) that may take orders of magnitude less time to solve than (\ref{eq:pi_B^* brute force}), as shown in Fig.~\ref{fig:timeComplexityOfExactEffectiveLossRate}.


In order to compute $\pi_B^*$ under the even-spacing case, we look closer at the packets lost on every path separately. Denote by $F_r$ and $D_r$ the number of FEC and data packets lost on path $P_r$, respectively (both before FEC recovery). Now we can rewrite the total number of lost FEC packets as $F=\sum_{r}F_r$ and the total number of lost data packets as $D=\sum_{r}D_r$.
This decomposition leads us to the following derivation of $\pi_B^*$:

\noindent \includegraphics[width=0.49\textwidth]{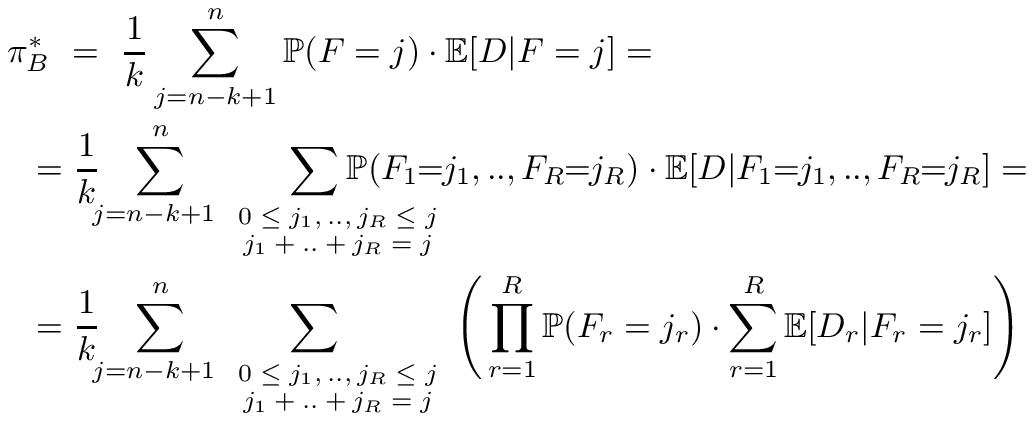}
\nopagebreak
\vspace{-0.9cm}
\nopagebreak
\begin{equation}\label{eq:E[D]}
{}
\end{equation}


According to Equation~(\ref{eq:E[D]}), in order to evaluate $\pi_B^*$, for every path $P_r$ separately we need to calculate two components: (i) the probability $\pr(F_r\!=\!j_r)$ that $j_r$ FEC packets are lost, and (ii) the expected number $\E[D_r | F_r\!=\!j_r]$ of lost data packets given that $j_r$ FEC packets were lost. We achieve this by applying an approach similar to the one used in~\cite{Frossard01} in the context of a single path FEC, as follows.

We consider a path $P_r$ and a set of all $n_r$ FEC packets sent on $P_r$ with equal packet interval $T_r$. Denote by $[{\textstyle {a \atop b}}]$ the event that any $b$ out of $a$ consecutive packets are lost.\footnote{The form of $[{\textstyle {a \atop b}}]$ is inspired by the similarity with the binomial coefficient.} We allow for concatenation of events, e.g., $G[{\textstyle {a \atop b}}]$ (resp., $[{\textstyle {a \atop b}}]B$) means that any $b$ out of a block of $a$ consecutive packets are lost and that this block is preceded by a good packet (resp., followed by a bad packet). We can now compute $\pr(F_r = j_r)$ by conditioning on the state of the first packet that conforms the packet loss stationary distribution:
\begin{eqnarray}
\nonumber && \hspace{-0.5cm} \pr(F_r = j_r) = \pr(G\ [{\textstyle {n_r-1 \atop j_r}}]) +\ \pr(B\ [{\textstyle { n_r-1 \atop j_r-1}}])\ =\\
\label{eq:Pr(F_r = j_r)}  && =\pi_G^{(r)}\cdot \pr([{\textstyle { n_r-1 \atop j_r}}]\ |\ G) +\ \pi_B^{(r)}\cdot \pr([{\textstyle {n_r-1 \atop j_r-1}}]\ |\ B),\qquad {}
\end{eqnarray}
where $\pr([{\textstyle {a \atop b}}]\ |q)$, $q\in\{G,B\}$, is the probability that any $b$ out of $a$ consecutive packets are lost given that this block is preceded by a packet in state $q$. Although no general closed form of $\pr([{\textstyle {a \atop b}}]\ |q)$ is known, it can be calculated by the recursive approach first proposed in \cite{Elliott1965} and extended e.g. in~\cite{Frossard01,Golubchik02}. We show in Appendix the details of this computation. It takes $\pi_B^{(r)}$, $1/\mu_B^{(r)}$ and $T_r$ as parameters, and directly uses the relations (\ref{eq:p^r_i,j(tau)}) above.

In order to find $\E[D_r | F_r\!=\!j_r]$, we first derive $\pr(D_r\!=\!i, F_r\!=\!j_r)$. Let us consider the $k_r$ data packets and the $n_r\!-\!k_r$ redundancy packets separately, and additionally condition on the state of the last data packet as follows.\smallskip

\noindent \includegraphics[width=0.49\textwidth]{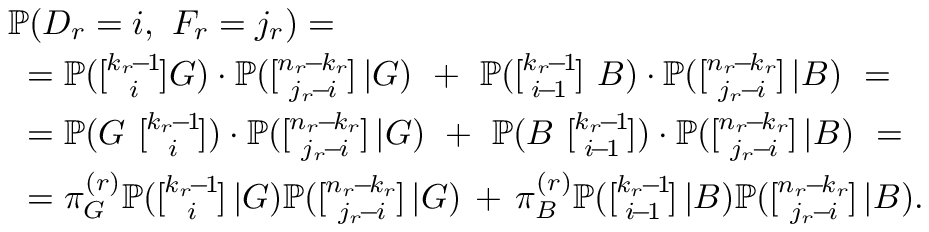}
%
%
The first equality uses the Markov property of the loss model:\smallskip

\noindent \includegraphics[width=0.49\textwidth]{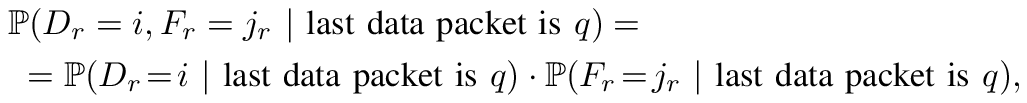}
where $q\in \{G,B\}$. Now it is easy to calculate $\E[D_r | F_r=j_r]$, because
\begin{eqnarray}
\label{eq:E[D_r | F_r=j_r]}  \E[D_r | F_r=j_r] &=& 
\sum_{i=0}^{k_r} i\cdot \frac{\pr(D_r=i, F_r = j_r)}{\pr(F_r = j_r)}.
\end{eqnarray}
We plug (\ref{eq:Pr(F_r = j_r)}) and (\ref{eq:E[D_r | F_r=j_r]}) into (\ref{eq:E[D]}) and obtain a complete formula for the effective loss rate $\pi_B^*$:

\noindent \includegraphics[width=0.49\textwidth]{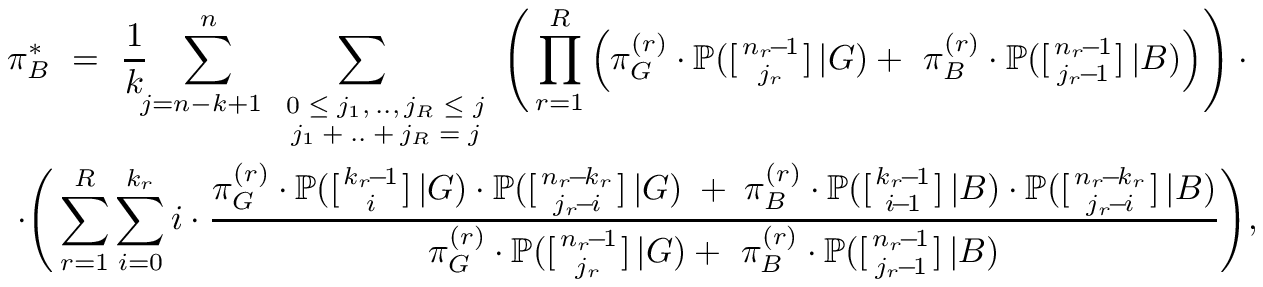}
\nopagebreak
\vspace{-0.7cm}
\nopagebreak
\begin{equation}\label{eq:BIG pi_B*}
{}
\end{equation}
\nopagebreak
where every term of type $\pr([{\textstyle {a \atop b}}]\ |G)$ or $\pr([{\textstyle {a \atop b}}]\ |B)$ is calculated through the set of recursive equations given in Appendix.\smallskip
\begin{figure}[!t]
    \psfrag{Y}[c][c]{\large $\frac{\widetilde{\E}[D_r | F_r=j_r]}{\E[D_r | F_r=j_r]}(\frac{\textrm{approx.}}{\textrm{exact}})$}
    \psfrag{j}[c][c]{j}
    \begin{center}\includegraphics[width=0.3\textwidth]{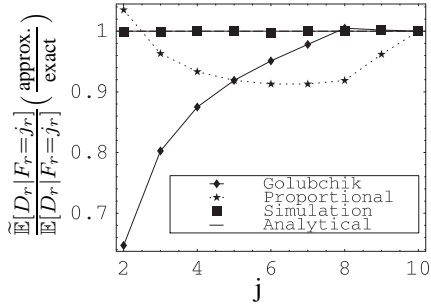}\end{center}
\caption{Approximations of $\E[D_r | F_r=j_r]$ normalized by the correct value given by (\ref{eq:E[D_r | F_r=j_r]}). Here $n_r=10$, $k_r=8$, $\pi_B^{(r)}=0.01$ and $1/\mu_B^{(r)}=2$. }
\label{fig:approximations}
\end{figure}

To the best of our knowledge, Equation~(\ref{eq:BIG pi_B*}) is the first exact solution of this model. Indeed, all previous works used some approximations of $\E[D_r | F_r=j_r]$. In~\cite{Golubchik02} the authors approximate $\E[D_r | F_r=j_r]$ by assuming that any configuration of $j$ losses among the $n$ FEC packets is equally likely; we call this approach `Golubchik'. In~\cite{Yu05,Yu05b} the authors use an intuitive linear formula, i.e., $\E(D_r| F_r = j_r)=\frac{k_r}{n_r}j_r$. Although not mentioned in the papers this is only an approximation that is exact only when $k_r,n_r\rightarrow \infty$; we refer to it as `Proportional'. We illustrate the differences between these approximations and the real values in Fig.~\ref{fig:approximations}.

\section{The design of the schedule~$\Sch$}\label{sec:Packet_scheduling}
In the previous section we derived an exact analytical formula for the effective loss rate $\pi_B^*$ under a given schedule~$\Sch$. Here we focus on the design of a good schedule that results in small~$\pi_B^*$.


Not all schedules are applicable in practice. Indeed, both (i) the maximal allowed FEC block transmission time $\tfec$ and (ii) the packet interval $T$ at the source impose important scheduling constraints. We say that a schedule is \emph{feasible} if all the three following conditions are satisfied:\smallskip

\noindent\textbf{C1}\ $\T(i)\geq (i-1)\cdot T$ for $1\!\leq\!i\!\leq\!k$,\quad i.e., no data packet is sent before it is generated at the source.\smallskip

\noindent\textbf{C2}\ $\T(i)\geq (k-1)\cdot T$ for $k\!<\!i\!\leq\!n$,\quad i.e., no redundancy packet is sent before all data packets have been generated (we need to collect all data packets in order to create the redundancy packets).\smallskip

\noindent\textbf{C3}\ $\T(i)+t_{\R(i)} \leq \tfec$\ for\ $1\!\leq\!i\!\leq\!n$,\quad i.e., all FEC packets should arrive at the destination before the deadline.\smallskip

%
%
We assume that the path rates $n_1,\dots,n_R$ are fixed. There are usually a variety of feasible schedules. Below we discuss two classes of schedules we use in this paper.  The first one, called \emph{Immediate}, reflects the state of the art, whereas the second one, \emph{Spread}, is our proposal.

\subsection{`Immediate' packet scheduling $\Sch^{imm}$ -  state of the art} \label{subsec:Immediate packet scheduling Sch^imm}
We denote by \emph{Immediate} the schedule $\Sch^{imm}=(\T^{imm},\R^{imm})$ that represents the approach used in \cite{Golubchik02,Nguyen03path,Vergetis05,Yu05b,Levy06,Li07SmartTunnel}. As the name suggests, Immediate sends the data packets as soon as they are generated, i.e., every time interval~$T$. The redundancy packets use the same spacing $T$. So in general
\begin{equation}\label{eq:Immediate_Times_in_Schedule}
    \T^{imm}(i)=(i-1)\cdot T \textrm{\quad for $1\leq i\leq n$.}
\end{equation}
This specifies \emph{when} the FEC packets are sent, but not on which path. A good and commonly used guideline for $\R^{imm}$ is to spread the packets on each path separately with (roughly) even spacing~\cite{Li07SmartTunnel}.
When the rates are equal, i.e., $n_1\!\!=\!\!n_2\!\!=\!\!\ldots\!\!=\!\!n_R$, then this boils down to a simple round-robin schedule applied in~\cite{Golubchik02,Vergetis05,Yu05b,Levy06}. In contrast, when the rates differ, a more elaborate approach should be used. For this purpose we adopt the credit-based technique proposed in~\cite{Li07SmartTunnel}, as follows. Each path is associated with a credit initially equal to 0. Before each FEC packet transmission the credit of every path $P_r$ is increased by $n_r/n$. Next, the path with the largest credit is selected to transmit this packet; the credit of this path is decreased by 1. This scheme is iterated until all $n$ FEC packets are sent.

The Immediate schedule can be interpreted as the following function:
$$\Sch^{imm}=Immediate(n_1\ldots n_R,\ T)$$


Two examples of Immediate schedules $\Sch^{imm}$ are given in Fig.~\ref{fig:schedulings}: (C) is a single-path schedule, i.e., with $n_1=6$ and $n_2=0$, whereas in (D) we use two paths and $n_1=n_2=3$.


\subsection{`Spread' packet scheduling $\Sch^{spr}$ - our proposal} \label{subsec:Spread packet scheduling}
Under Immediate, all packets are sent as soon as they are generated, i.e., according to (\ref{eq:Immediate_Times_in_Schedule}). Instead, we propose to \emph{spread the packets evenly in all the available time on each path}. We call this schedule \emph{Spread} $\Sch^{spr}=(\T^{spr},\R^{spr})$. Compared with Immediate, Spread additionally takes the path propagation times $t_1\ldots t_R$ and the maximal FEC block delay $\tfec^{spr}$ as parameters, i.e.,
$$\Sch^{spr}=Spread(n_1\ldots n_R,\ T,\ t_1\ldots t_R,\ \tfec^{spr}).$$


The design of Spread is not straightforward. Indeed, as the~$k$ data packets are generated at the source with spacing~$T$, the paths are  inter-dependent, which may easily lead to the violation of the constraint C1.
For example, if we schedule packet~1 on $P_1$ at time $\T(1)=0$ (and $k>1$), then no other packet on any path can be scheduled before time $t=T$.

We guarantee the feasibility of Spread as follows.
First, we order the paths according to their rates, starting from the path with the highest rate. (When two paths have the same rate, we take the one with a higher path propagation time first.) We consider the paths one by one following this order. For each path $P_r$ we spread the packets evenly on time interval $[t^{(r)}, \tfec^{spr}\!-\!t_r]$, where $t^{(r)}$ takes the smallest possible value that satisfies the feasibility condition. (The value of $t^{(r)}$ usually grows with the number of paths processed.) We iterate this algorithm until all paths have been scheduled.


We present two examples of Spread schedules $\Sch^{spr}$ in Fig.~\ref{fig:schedulings}. We use  $\tfec^{spr}=170ms$ and two different sets of rates: $n_1\!=\!n_2\!=\!3$ in (E) and $n_1\!=\!4$, $n_2\!=\!2$ in (F).

Spread is very effective. Indeed, we can prove that
\begin{theorem}\label{th:Optimality of Spread for FEC(n,1)}
The Spread schedule is optimal for the repetition code FEC$(n,1)$.
\end{theorem}

\noindent \textbf{Proof of Theorem~\ref{th:Optimality of Spread for FEC(n,1)}}
Under FEC$(n,1)$ every data packet is replicated and sent in $n$ copies; the reception of at least one such copy leads to a success.  As there is only one data packet, all the redundancy packets (i.e., the duplicates of the data packet) can be generated already at time $t\!=\!0$. This eliminates all the time dependencies between the paths. Therefore, every path $P_r$ separately must maximize the probability of at least one successful transmission. It is achieved by \emph{even} spreading on the maximal allowed time interval $[0, \tfec^{spr}\!-\!t_r]$. (The proof for the under repetition code on a single path can be found in~\cite{Bolot99}.) This, in turn, is exactly what Spread returns under FEC$(n,1)$.


\vspace{-0.5cm}\begin{flushright}$\blacksquare$\end{flushright}

%

Spread builds on even packet spreading  - a simple and widely accepted guideline that is often thought of as leading to the optimal solution. Indeed, its optimality was proven for some particular cases~\cite{Bolot99}. But, surprisingly, this is not a general result. Consider for example FEC(4,3) on a single path (i.e., $\R=(1,1,1,1)$) with loss rate $\pi_B^{(1)}=1\%$ and average loss burst length $1/\mu_B^{(1)}=5ms$, and available time interval equal to 15ms. The even spreading schedule $\Sch_1=((0,5,10,15),\R)$ yields $\pi_B^*=0.53\%$. But the optimal schedule (found with optimization tools of Mathematica~\cite{WWW_Mathematica}) is $\Sch_1=((0,7.16,12.51,15),\R)$ and yields $\pi_B^*=0.50\%$.

This means that Spread does \emph{not} guarantee optimality in the general FEC$(n,k)$ case. However, we show later in simulations that it usually leads to almost-optimal solutions and is thus an effective and practical rule of thumb.

\subsection{Comparison of $\Sch^{imm}\!$ and $\Sch^{spr}$: Optimal schedules $\Sch^{imm}_{opt}$ and $\Sch^{spr}_{opt}$, and loss rate improvement $\gamma$.}\label{subsec:gamma}
It was shown in previous studies that Immediate multipath is better than a single path communication. The main point we make here is that once we allow for multipath, the Spread schedule $\Sch^{spr}$ that we propose in this paper is significantly better than the Immediate schedule $\Sch^{imm}$ representing the state of the art.

In order to demonstrate this, we compare the performance of $\Sch^{imm}$ and $\Sch^{spr}$ in terms of their effective loss rates. What rates $n_1\ldots n_R$ and what FEC block transmission time $\tfec$ should we use to make this comparison meaningful and fair? We should allow Immediate and Spread to optimize their rates $n_1\ldots n_R$ independently, given that they impose identical FEC block transmission times $\tfec^{imm}=\tfec^{spr}$. More precisely, we assume that the FEC parameters $n$ and $k$ are fixed, and we proceed in two steps. First, we optimize the rates $n_1\ldots n_R$ of Immediate, such that the 
effective loss rate $\pi_B^*$ is minimized. It results in the optimal Immediate schedule $\Sch^{imm}_{opt}$. This, in turn, specifies $\tfec^{imm}$ as shown in~(\ref{eq:t_fec}). In the second step, we set $\tfec^{spr}=\tfec^{imm}$ and optimize the rates $n_1\ldots n_R$ of Spread, resulting in the optimal Spread schedule $\Sch^{spr}_{opt}$.\footnote{Note that $\Sch^{imm}_{opt}$ and $\Sch^{spr}_{opt}$ are optimal subject to their construction constraints presented in \ref{subsec:Immediate packet scheduling Sch^imm} and \ref{subsec:Spread packet scheduling}, respectively.}

Finally, we define the \emph{relative effective loss rate improvement} $\gamma$ as the relative gain in $\pi_B^*$ due to the usage of optimal Spread instead of optimal Immediate, i.e.,
\begin{equation}\label{eq:gamma}
     \gamma = \frac{\pi_B^*(\Sch^{imm}_{opt})}{\pi_B^*(\Sch^{spr}_{opt})}.
\end{equation}
The metric $\gamma$ can be precisely evaluated by formulas~(\ref{eq:pi_B^* brute force}) and (\ref{eq:BIG pi_B*}). The values of $\gamma$ can be easily interpreted; for example, $\gamma>1$ means that Spread performs better than Immediate.

\subsection{Capacity constraints}
So far we have considered the case where every path $P_r$ can be assigned with any rate 0$\leq n_r\leq n$. In practice, however, $P_r$ may have a relatively limited capacity, which would impose a direct constraint on $n_r$. Fortunately, integrating these constraints in our model is straightforward. Indeed, it is enough to respect them when computing the rates $n_1\ldots n_R$ in $\Sch^{spr}_{opt}$ and $\Sch^{spr}_{imm}$ in~\ref{subsec:gamma}.

\section{Performance evaluation}\label{sec:Performance evaluation}
In this section we evaluate our approach first in simulations
and next on real-life traces.

\subsection{Simulation results}\label{subsec: FEC Simulation results}
The goal of simulations is twofold. First, we verify the correctness of our analytical results. Second, we can test our idea in a fully controlled environment and study the effect of various parameters on the results.


\smallskip
\subsubsection{Default values of parameters}
If not stated otherwise, in our simulations we use the following default values. The data packets are generated at the source with interval $T=5ms$. Next they are encoded by systematic FEC$(10,8)$ and sent over $R$ independent paths.
For the sake of simplicity we speak mainly of systems with $R=2$ paths: $P_1$ and $P_2$. It allows us to describe the path propagation time differences by a single parameter $\Delta t= t_2\!-\!t_1$ that takes the default value $\Delta t=100ms$. Finally, the paths $P_1$ and $P_2$ have the same average failure rate $\pi_1=0.01$ and the average loss burst length equal to $1/\mu_1=10ms$.

\smallskip
\subsubsection{The effective loss rate $\pi^*_B$ as a function of $\Delta t$}

\begin{figure}[!t]
    \psfrag{LOSS_PR}[c][]{Effective loss rate $\pi^*_B$}
    \psfrag{LOSS_PR2}[c][]{\small $\pi^*_B$}
    \psfrag{n1}[c][]{\small $n_1$ - rate on $P_1$}
    \psfrag{dt50}[c][c][0.7]{\small $\Delta t=50ms$}
    \psfrag{delta}[c][b]{Propagation time difference $\Delta t$}
    \psfrag{single}[r][c][0.8]{\small Single path Immediate $\Sch^{imm}_{(10,0)}$}
    \psfrag{imm}[r][c][0.8]{\small\hspace{-1cm} Immediate optimal $\Sch^{imm}_{opt}\!\!=\!\Sch^{imm}_{(5,5)}$ (state of the art)}
    \psfrag{sprOpt}[c][b][0.8][-18]{\small Spread optimal $\Sch^{spr}_{opt}$}
    \psfrag{spr55}[r][b][0.8]{\small Spread $\Sch^{spr}_{(5,5)}$}
    \psfrag{circle}[l][]{\small simulation}
    \psfrag{lines}[l][]{\small analytical}
    \psfrag{optim}[l][]{\small optimal}
    \begin{center}\includegraphics[width=0.48\textwidth]{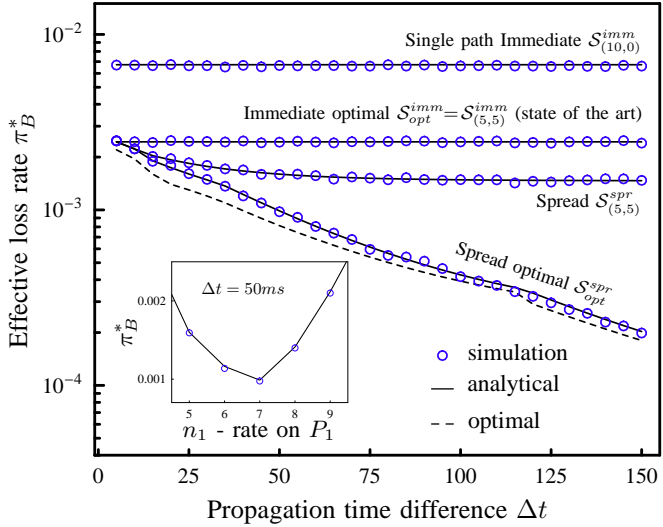}\end{center}
\caption{The effective loss rate $\pi^*_B$ as a function of path propagation time difference $\Delta t$. We use $FEC(10,8)$ on two independent paths, $P_1$ and $P_2$, with data packet spacing $T=5$ at the source. The losses on $P_1$ and $P_2$ are modeled by continuous time Gilbert model with the same average failure rate $\pi_B=0.01$ and the average burst length equal $1/\mu_B=10ms$. Four schedules are used:\quad $\bullet$~$\Sch^{imm}_{(10,0)}$ - all packets are sent on a single path $P_1$ with interval $T$, \quad $\bullet$~$\Sch^{imm}_{(5,5)}$ - Immediate with optimal rates $n_1\!=\!n_2\!=\!5$, \quad $\bullet$~$\Sch^{spr}_{(5,5)}$ - Spread with $n_1\!=\!n_2\!=\!5$,  \quad $\bullet$~$\Sch^{spr}_{opt}$ - Spread with the rates $n_1,n_2$ chosen optimally based on the value of $\Delta t$. \quad Additionally, the dashed curve shows the effective loss rate of the optimal schedule, where packets are not restricted to even spacing on each path, as described in Section~\ref{sec:Packet_scheduling}. The optimal schedule was found with numerical optimization tools of Mathematica~\cite{WWW_Mathematica}. \qquad \textbf{Inset:} $\pi^*_B$ as a function of rate $n_1$ on path~$P_1$ for $\Delta t=50ms$ under Spread. \quad In both figures the plain lines are the theoretical values according to
formula~(\ref{eq:BIG pi_B*}), whereas the circles are the results obtained in a simulation of the model. The size of confidence intervals (not shown) is comparable with the size of the circles.}\label{fig:various_schedules}
\end{figure}

In Fig.~\ref{fig:various_schedules} we plot the effective loss rate $\pi^*_B$ as a function of $\Delta t$ for four different schedules. Our first observation is that the results obtained in a simulation of the model (circles) fit precisely the analytical curves (plain lines).

Next, we compare the performance of various schedules. As the loss properties of the two paths are identical, the previous techniques described in~\cite{Golubchik02,Nguyen03path,Li07SmartTunnel} split the FEC packets equally between $P_1$ and $P_2$. This results in the optimal Immediate schedule $\Sch^{imm}_{opt}=\Sch^{imm}_{(5,5)}$, i.e., with $n_1\!=\!n_2\!=\!5$. As this schedule uses multipath transmission, it is not surprising that $\Sch^{imm}_{opt}$ significantly outperforms the single path Immediate schedule $\Sch^{imm}_{(10,0)}$. Note also that, by construction, $\Delta t$ does not affect the performance of any of them.

In contrast, in Spread $\Sch^{spr}_{(5,5)}$ we use the same rates as in $\Sch^{imm}_{opt}$, but we spread the packets uniformly within the time budget $\tfec^{imm}$ set by $\Sch^{imm}_{(5,5)}$ (similar schedule is shown in Fig.~\ref{fig:schedulings}E).
It results in a further decrease of the effective loss rate $\pi^*_B$. This difference moderately grows with $\Delta t$. However, for larger $\Delta t$ the rates $(5,5)$ become suboptimal under Spread. For instance, in the inset in Fig.~\ref{fig:various_schedules} we show the performance of Spread under various rate configurations $(n_1, n\!-\!n_1)$; the minimum is reached for $(7,3)$. As descried in~\ref{subsec:gamma}, allowing for this rate optimization leads to the optimal Spread schedule $\Sch^{spr}_{opt}$. Its advantage over $\Sch^{imm}_{(5,5)}$ grows roughly exponentially with $\Delta t$.

Finally, we observe that the performance of the optimal Spread schedule $\Sch^{spr}_{opt}$ is very close to the global optimum (dashed curve) where packets are not necessarily evenly-spaced, as described in Section~\ref{sec:Packet_scheduling}. This confirms the usefulness of the even-spread guideline that we follow in Spread.

\smallskip
\subsubsection{Loss rate improvement $\gamma$ as a function of various parameters}

\begin{figure*}[!t]
    \psfrag{A}[c][]{(A)}
    \psfrag{B}[c][]{(B)}
    \psfrag{C}[c][]{(C)}
    \psfrag{D}[c][]{(D)}
    \psfrag{gamma}[c][]{\small{Improvement} $\gamma$}
    \psfrag{gam}[c][]{$\gamma$}
    \psfrag{dT}[c][c][1] {$\Delta t$}
    \psfrag{T}[c][c][1]{$T$}
    \psfrag{pi}[c][c][1]{$\pi_B^{(2)}$}
    \psfrag{pi1}[c][c][0.7]{\small $\pi_B^{(1)}\!=\!0.01$}
    \psfrag{N}[c][c][1]{$n$}
    \psfrag{ratio}[c][c][0.7][-8]{${}\ \pi_B^{*}(\Sch^{imm}_{(10,0)})/\pi_B^*(\Sch^{imm}_{opt})$}
    \begin{center}\includegraphics[width=1.\textwidth]{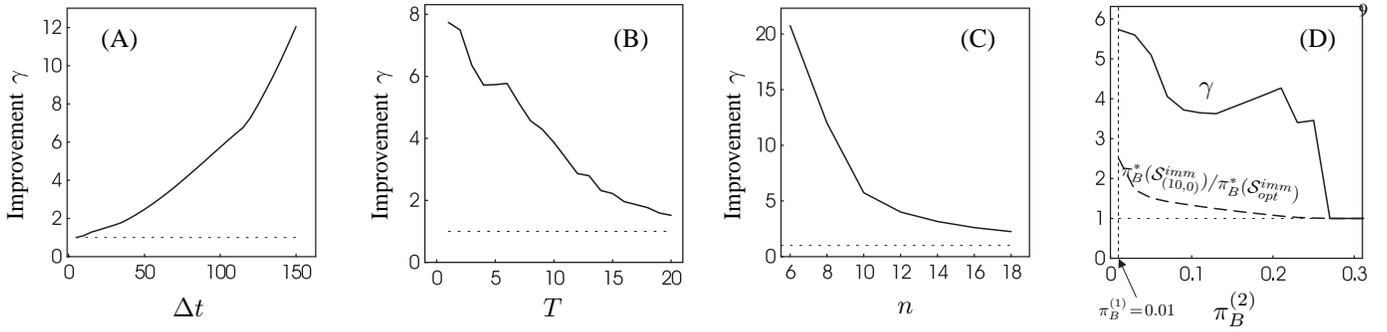}\end{center}
\caption{Relative loss rate improvement $\gamma$ due to usage of Spread instead of Immediate as a function of four parameters: (A)~path propagation time difference $\Delta t$, \ (B)~packet generation interval $T$ at the source, \ (C)~the size $n$ of a FEC block, \ (D)~loss rate $\pi_B^{(2)}$ of path $P_2$. We consider a system with $R=2$ paths and the following default parameters: FEC(10,8), $\Delta t=100ms$, $T=5ms$, $\pi_B^{(r)}=1\%$, $1/\mu_B^{(r)}=10ms$, $k=n-2$. All results shown here are analytical.\quad  The irregular shapes of the curves in this and other figures are expected, because the computation of $\gamma$ involves the rates optimization (see~\ref{subsec:gamma}). For instance, in figure (D), going from left to right, the optimal Immediate and Spread rates $(n_1,n_2)$ change gradually (and separately) from $(5,5)$ to $(10,0)$; every such rate transition may introduce irregularities in the shape of the curves.}\label{fig:gamma}
\end{figure*}

Clearly, there are many parameters that affect the performance of the schedules. We study the effect of some of them on the relative loss rate improvement $\gamma$ in Fig.~\ref{fig:gamma}.

First, plot (A) confirms that the advantage of Spread over Immediate grows with the path propagation time difference~$\Delta t$.

Second, with growing packet interval $T$ at the source, the fixed $\Delta t$ becomes a smaller fraction of the entire FEC block transmission time $\tfec$. As a consequence, there is relatively less to exploit and $\gamma$ drops with $T$, see plot~(B). A similar phenomenon can be observed in plot (C), where $\tfec$ grows due to an increase of the number $n$ of FEC packets.

Finally, in Fig.~\ref{fig:gamma}D we vary the loss rate $\pi^{(2)}_B$ of path $P_2$.
The difference between path loss rates is a crucial parameter affecting the performance gain of Immediate multipath over the single path transmission. Indeed, if out of two paths one is very lossy and the other one is very good, then the optimal Immediate multipath schedule $\Sch^{imm}_{opt}$ uses mainly (or only) the better path, which substantially limits the gain of multipath~\cite{Nguyen03path,Vergetis05}. This is illustrated in plot (D) by the dashed curve; the ratio
$\pi_B^{*}(\Sch^{imm}_{(10,0)})/\pi_B^*(\Sch^{imm}_{opt})$
is the largest when the paths have identical loss properties, and quickly diminishes with growing difference between $\pi^{(1)}_B$ and $\pi^{(2)}_B$.

We could expect a similar diminishing effect for the advantage $\gamma=\pi_B^*(\Sch^{imm}_{opt})/\pi_B^*(\Sch^{spr}_{opt})$ of Spread over Immediate. Surprisingly, this is not the case; $\gamma$ remains relatively stable ($3\!<\!\gamma\!<\!6$) for a wide range of values of $\pi^{(2)}_B$. For $\pi^{(2)}_B\approx 0.25$ the path $P_2$ becomes too lossy, and both Immediate and Spread send all packets on $P_1$ only and thus become equivalent.

\smallskip
\subsubsection{Minimizing $\tfec$ - decreasing delays and fighting jitter}

\begin{figure}[!t]
    \psfrag{delay_gain}[c][t]{$\tfec^{imm}-\tfec^{spr}\ \ [ms]$}
    \psfrag{dT}[c][b]{Propagation time difference $\Delta t$ $[ms]$}
    \begin{center}\includegraphics[width=0.3\textwidth]{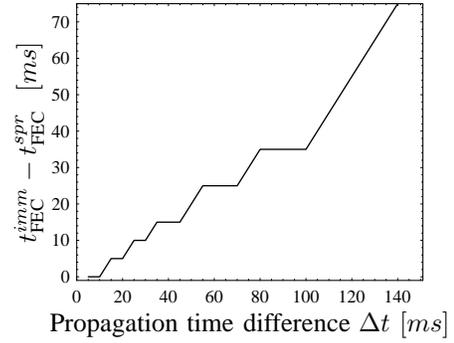}\end{center}
\caption{The gain in FEC block transmission time $\tfec$ by the usage of Spread instead of Immediate. Parameters: FEC(10,8), $\pi^{(1)}_B=\pi^{(2)}_B=0.01$, $1/\mu_B^{(1)}\!=\!1/\mu_B^{(2)}\!=\!10ms$, $T=5ms$. For these parameters, the optimal Immediate rates are $n_1\!=\!n_2\!=\!5$, which results in the effective loss rate $\pi_B^*(\Sch^{imm}_{(5,5)})\!\!=\!0.24\%$. For Spread we choose the minimal FEC block transmission time $\tfec^{spr}$ such that  $\pi_B^*(\Sch^{spr}_{opt}) \leq \pi_B^*(\Sch^{imm}_{(5,5)})$.}\label{fig:delay_gain}
\end{figure}

So far we used Spread to minimize the effective loss rate $\pi_B^*$ while keeping the FEC block transmission time $\tfec^{spr}$ not larger than that of Immediate schedule $\tfec^{imm}$. Let us now reverse the problem: Let us minimize the FEC block transmission time $\tfec^{spr}$ of Spread, while keeping its effective loss rate not larger than that of Immediate, i.e., subject to $\pi_B^*(\Sch^{spr}_{opt}) \leq \pi_B^*(\Sch^{imm}_{opt})$.

We plot the results in Fig.~\ref{fig:delay_gain}. The gain $\tfec^{imm}\!-\tfec^{spr}$ in FEC block transmission time is significant and grows roughly linearly with $\Delta t$, as $\tfec^{imm}\!-\tfec^{spr}\simeq \Delta t/2$. The reduction of $\tfec$ brings obvious advantages to delay-constrained applications using the multipath FEC system. First, the effective end-to-end delays get smaller which allows us to reduce the playout time at the destination, keeping the same level of the effective loss rate.

Another important interpretation is related to the delay \emph{jitter}, i.e., variations of path propagation times. Indeed, in this work, as in most previous works on multipath FEC, we consider the path propagation times constant and focus on (correlated) packet losses only. However, as Spread results in a smaller delay $\tfec$, it also leaves more space to accommodate potential jitter, naturally making Spread more robust to jitter than Immediate.

%

\smallskip
\subsubsection{Other FEC parameters $n,k$}
So far we assumed that Immediate and Spread use the same general FEC parameters $n$ and $k$; only the rates on particular paths could be optimized. However, in some cases the optimal choice of $n$ and $k$ under Spread may differ from that of Immediate, given the same redundancy $k/n$.

For example, according to our additional simulations (not shown here), for the setting in Fig.~\ref{fig:various_schedules} and $\Delta t>220ms$, the Spread schedule using FEC(15,12) would outperform Spread with FEC(10,8). Similarly, FEC(12,6) would be better for Spread than FEC(10,5) for $\Delta t>140ms$. Note, however, that this phenomenon can be observed only for relatively large values of $\Delta t$ that rarely occur in reality.\smallskip

To conclude,
the loss rate improvement $\gamma$ can be manyfold, but its exact value strongly depends on various parameters. First, the advantage of Spread over Immediate grows with path propagation time differences $\Delta t$, but drops with the data packet generation interval $T$ and FEC block size $n$. Second, the optimal Immediate rates are not always optimal under Spread; usually optimal Spread sends more packets on faster paths. Third, although the advantage of Immediate over a single path transmission quickly diminishes with growing differences between the loss rates of the paths, the advantage of Spread over Immediate is relatively stable. Finally, Spread can also achieve FEC block transmission times much smaller than Immediate, still guaranteing the same or better performance in terms of the effective loss rate. This results not only in smaller effective delays, but also in a higher robustness to delay jitter.

\subsection{Trace-driven PlanetLab evaluation}
In the previous section we presented analytical and simulation results where the packet losses were modeled by the Continuous Time Gilbert Model. As any model, it is only an approximation of reality. In this section we feed our simulations with real-life packet loss traces collected in Internet experiments.

\subsubsection{Data sets}

The traces come from two different PlanetLab (PL)~\cite{WWW_PlanetLab} experiments. On every path the packets are sent with time-interval $T$, i.e., with the generation rate at the source. Every trace is a sequence composed of symbols $G$ (packet not lost) and $B$ (packet lost).

Every time-constrained experiment on PlanetLab should be designed and interpreted carefully. This is because at any point in time most of PlanetLab nodes are overloaded. Not only their CPU utilization is at 100\%, but more importantly the queueing delays experienced by the running processes can be very significant - even up to several seconds between two consecutive accesses to CPU. This results in incorrect propagation time measurements and packet dropping due to incoming buffer overflow at the destination~\cite{Amir06,Sommers07}. Moreover, the situation changes dynamically.
We minimize the effects of these problems by introducing periodic pauses in packet generation and avoiding the highly loaded PlanetLab nodes.

We use the following two data sets.

\paragraph{`Relays' - PlanetLab with relays}
In this experiment every trace is collected on a two-hop overlay path between three PlanetLab nodes: source, relay and destination. The UDP packets at the source are generated every $T=5ms$ and sent immediately to the relay that forwards them to the destination. After every one-second-long packet generation period we introduce one second of idle time in order to avoid dropping packets at PlanetLab hosts when the probing traffic is too bursty. We collected more than 5'000 traces, each covering 100 seconds of packet generation time.

In order to further reduce the effect of overloaded PL nodes on the results, for every experiment separately we select the source, relay and destination randomly from 50 currently least loaded PL nodes. As the load estimate we use the number of processes queueing for the CPU and I/O devices; it can be obtained by parsing the file \verb|/proc/stat| that stores the information about kernel activity.


\paragraph{`Web sites' - PlanetLab to popular web sites}
This data set consists of 2'839 traces used in~\cite{Li07SmartTunnel}. They were collected by sending 16-byte ICMP echo packets from 57 PlanetLab hosts to 55 popular web sites selected from~\cite{WWW_100hotsites}. Next, the ICMP echo-reply packets were captured by Tcpdump, resulting in traces where packets travel from a PlanetLab node to a web site and back to the original PlanetLab node. The packets were sent every $T=2ms$. As above, every one-second packet generation time was followed by one-second idle time.  Each measurement lasted at least 800 seconds. As in~\cite{Li07SmartTunnel}, we split it into 40-second long intervals that we call chunks.

Despite the measures we took, in both data sets we find traces with numerous long (100ms and more) blocks of consecutive losses. As this is not caused by network losses, but rather by buffer overflow at the nodes due to CPU queueing, we exclude these traces from our simulations.

\subsubsection{Simulation setting}
In a simulation of a $R$-path scenario we use $R$ traces (one per path) randomly chosen from the pool of all available traces. Thus, by construction, the $R$ traces are independent (typically generated at different times and places in the Internet). For the sake of simplicity, we restrict the presentation to the case $R=2$.

Our basic metric is loss rate improvement $\gamma$. As described in Section~\ref{subsec:gamma}, it optimizes the rates of Immediate and Spread. This optimization is based on the observed traces. One approach to do this is to infer for every path its loss rate $\pi_B^{(r)}$ and the average loss burst length $1/\mu_B^{(r)}$, feed them into the model and optimize the rates as in section~\ref{subsec: FEC Simulation results}. However, this technique has two drawbacks: it introduces errors when measuring the path properties, and assumes a particular packet loss model. We avoid these problems by working directly on the traces - the optimal rates in $\Sch^{imm}_{opt}$ and $\Sch^{spr}_{opt}$ are those that perform best on a given chunk.

We present two types of results. In \emph{Oracle} we choose the optimal rates for the currently evaluated chunk. In contrast, in \emph{Prediction} we use the optimal rates of the preceding chunk to evaluate the current chunk. Thus Oracle shows the best achievable results for Immediate and Spread with no prediction errors, whereas Prediction is a practical implementation.


\subsubsection{Results}
\begin{figure}[!t]
    \psfrag{cdf}[c][]{CDF}
    \psfrag{improvement}[c][b]{\small Loss rate improvement $\gamma$}
    \psfrag{my}[l][]{`Relays'}
    \psfrag{st}[l][]{`Web sites'}
    \psfrag{t1}[c][c][1][10]{\footnotesize $\Delta t =10ms$}
    \psfrag{t5}[c][c][1][10]{\footnotesize $\Delta t =50ms$}
    \includegraphics[width=0.5\textwidth]{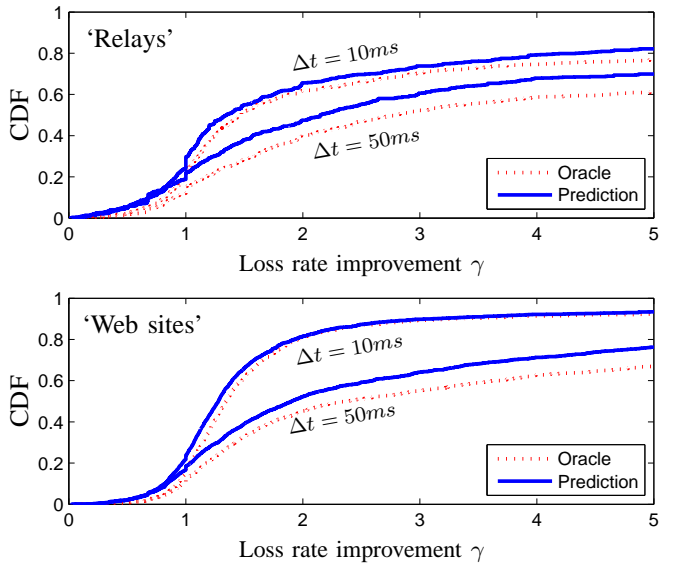}
\caption{The effective loss rate improvement $\gamma$ (by using Spread instead of Immediate) in trace-driven simulations under FEC$(10,8)$. We use $R=2$ independent paths with real-life loss traces; their propagation times differ by $\Delta t$. We consider two data sets: `Relays' (top) and `Web sites' (bottom).}\label{fig:trace_driven}
\end{figure}

In Fig.~\ref{fig:trace_driven} we present the results for FEC$(10,8)$. The figure presents the cumulative distribution of the relative loss rate improvement $\gamma$ for $\Delta t\!=\!10ms$ and $\Delta t\!=\!50ms$. We consider the cases where optimal Immediate uses both paths (i.e., $n_1\!\neq\! 0$ and $n_1\!\neq\! 10$) and there is a space for improvement (i.e., $\pi_B^*(\Sch^{imm}_{opt})>0$). In about 90\% of cases we observe an advantage of Spread over Immediate. For instance, for both data sets under Oracle with $\Delta t\!=\!50ms$, in 50\% of cases the loss rate drops by a factor of 3 or more when we use Spread instead of Immediate. For smaller $\Delta t$ the advantage is less pronounced, which is in agreement with the results presented in the previous section.

Surprisingly, in roughly 10\% of cases Spread performs slightly worse than Immediate. A possible explanation is that in some traces we can find loss patterns that are periodic, presumably due to other applications running on PlanetLab nodes. If such an unnatural loss pattern gets aligned with the packets scheduled by Spread on one or more paths, then the performance of Spread may drop below Immediate.

Finally, we find our simple prediction method satisfactory, as the Prediction curve is always close to Oracle.

\section{Related work}\label{sec:Related work}


The performance of FEC on a \emph{single} path with correlated loss failures was studied e.g., in~\cite{Bolot99,Frossard01,Jiang02}. One common conclusion is that the FEC efficiency drops with growing burstiness of packet losses.

\emph{Multipath} transmission as a way of de-correlating the packet losses and increasing the performance of FEC was first proposed in~\cite{Maxemchuk75}. It got more attention recently,
e.g., in~\cite{Golubchik02,Nguyen03path,Yu05b,Vergetis05,Levy06,Li07SmartTunnel}.
Multipath was also studied in the context of Multiple Description Coding~\cite{Apostolopoulos01}.

In~\cite{Nguyen03path} the authors study a multipath FEC system by simulations only, on artificially generated graphs. They also give a heuristic to select from a number of candidate paths a set of highly disjoint paths with relatively small propagation delays.

%

There are a number of approaches to evaluate \emph{analytically} the performance of multipath FEC with independent paths and bursty path losses. For instance, \cite{Golubchik02,Yu05b,Vergetis05} and \cite{Li07SmartTunnel} give four different derivations of the effective loss rate $\pi^*_B$ (or related metrics) in such a setting. However, in all four cases the resulting formula is only an \emph{approximation} of the complete solution due to (sometimes very significant) model simplifications. First, \cite{Yu05b}\cite{Vergetis05} use a discrete Gilbert model. Thus two consecutive packets on one path are equally correlated irrespectively of the time intervals between them, which makes the models inherently unable to capture any aspects of varying packet spacing. \cite{Li07SmartTunnel}~also uses the a discrete Gilbert model, but adapts the transition matrix appropriately. Second approximation comes when computing the number of lost data packets given that a FEC block cannot be entirely recovered: \cite{Golubchik02} and \cite{Yu05b}~use approximations described at the end of section~\ref{subsec:Performance of FEC on independent paths}, \cite{Vergetis05}~simplifies the model by assuming that in such a case all data packets are lost, and~\cite{Li07SmartTunnel} assumes that the numbers of lost data packets and redundancy packets are not correlated.
Third, \cite{Yu05b}~considers only a scenario with identical loss statistics on every path. Finally, \cite{Li07SmartTunnel} assumes a large number of active paths $R\gg 1$ and small individual path rates $n_r\ll n$. This allows the authors to apply the central limit theorem and approximate the joint distribution of the number of lost data and redundancy packets by a bivariate normal distribution.

To the best of our knowledge, we are the first ones to give an exact analytical formula for the effective loss rate $\pi^*_B$ of FEC protection scheme on multiple independent paths with path losses modeled by the Gilbert model.


As in most other approaches, we assume that the background cross-traffic is much larger than our own, and thus the load we impose on a path does not affect its loss statistics. Scenarios where this assumption does not hold were studied in~\cite{Yu05} in the context of a single path FEC, and in~\cite{Chow05} for multipath FEC.

As in~\cite{Golubchik02,Chow05,Levy06,Li07SmartTunnel} we assume the paths to be independent. This can be achieved by detecting correlated paths in end-to-end measurements~\cite{Rubenstein02} and treating them as one. 
Another approach is to find paths that are IP link disjoint, which should be possible if the site is multi-homed. Finally, even if all the available paths are to some extent correlated we can still get some performance benefits~\cite{Nguyen03path,Ribeiro05,Yu05b,JurcaFTMM:2006}, though limited~\cite{Andersen03,Tao04}.




Finally and \emph{most importantly}, to the best of our knowledge no attempt has been made to exploit the path propagation time differences in multipath FEC. Indeed, all the works listed above use some variant of the Immediate schedule, where packets are sent as soon as they arrive at the source.  In contrast, in this paper we have proposed the Spread schedule that exploits these propagation time differences and significantly improves the performance.
%

%

\section{Conclusion}

In this paper we started from the observation that the propagation times on multiple paths between a pair of nodes may significantly differ. We proposed to exploit these differences in the context of delay-constrained multipath systems using FEC, by applying the Spread schedule. We have evaluated our solution by a precise analytical approach, and simulations based on both the model and real-life Internet traces. Our studies show that Spread substantially outperforms previous solutions. It achieves a several-fold improvement (reduction) of the effective loss rate. Or conversely, keeping the same level of effective loss rate Spread significantly decreases the FEC block transmission time, which limits the observed delays and helps fighting the delay jitter.



\section{Acknowledgements}
We would like to thank the authors of~\cite{Li07SmartTunnel} for sharing their loss traces with us, and Patrick Thiran and Dan Jurca for helpful comments. This work is financially supported by grant ManCom 2110 of the Hasler Foundation, Bern, Switzerland.

\section{Appendix}
\subsection{Recursive equations}\label{appendix:Recursive equations}
Here we derive the probability $\pr([{\textstyle {a \atop b}}]\ |q)$ that any $b$ out of $a$ consecutive packets sent on a path $P_r$ (with packet interval $T_r$) are lost given that this block is preceded by a packet in state $q\in\{G,B\}$. Although no general closed form of $\pr([{\textstyle {a \atop b}}]\ |q)$ is known, it can be calculated by the recursive approach first proposed in \cite{Elliott1965} and extended e.g. in~\cite{Frossard01,Golubchik02}. Indeed,
\begin{eqnarray*}
  \pr([{\textstyle {a \atop b}}]\ |B) &=& R(b+1,a+1) \\
  \pr([{\textstyle {a \atop b}}]\ |G) &=& S(b+1,b-a+1),
\end{eqnarray*}
where functions $R(m,n)$ and $S(m,n)$ can be calculated as follows~\cite{Frossard01}:
\begin{eqnarray*}
R(m,n)\!\!\!&\!=\!&\!\!\! \left\{\!\! \begin{array}{ll}
    P(n) & \!\!\textrm{for $m\!=\!1$ and $n\!\geq\! 1$}\\
    \sum_{i=1}^{n\!-\!m\!+\!1} p(i) R(m\!\!-\!\!1,n\!\!-\!\!i)& \!\!\textrm{for $2\leq m\leq n$}
    \end{array} \right. \\
S(m,n)\!\!\!&\!=\!&\!\!\! \left\{\!\! \begin{array}{ll}
    Q(n) & \!\!\textrm{for $m\!=\!1$ and $n\!\geq\! 1$}\\
    \sum_{i=1}^{n\!-\!m\!+\!1} q(i) S(m\!\!-\!\!1,n\!\!-\!\!i)& \!\!\textrm{for $2\leq m\leq n$}
    \end{array} \right.
\end{eqnarray*}
where
\begin{eqnarray*}
p(i) &=& \left\{ \begin{array}{ll}
    1-q & \textrm{ if $i=1$}\\
    q(1-p)^{i-2}p& \textrm{ otherwise}
    \end{array} \right. \\
P(i) &=& \left\{ \begin{array}{ll}
    1 & \textrm{ if $i=1$}\\
    q(1-p)^{i-2}& \textrm{ otherwise}
    \end{array} \right. \\
q(i) &=& \left\{ \begin{array}{ll}
    1-p & \textrm{ if $i=1$}\\
    p(1-q)^{i-2}q& \textrm{ otherwise}
    \end{array} \right. \\
Q(i) &=& \left\{ \begin{array}{ll}
    1 & \textrm{ if $i=1$}\\
    p(1-q)^{i-2}& \textrm{ otherwise}
    \end{array} \right.\\
p &=& p^{(r)}_{G,B}(T_r) \textrm{\qquad - given by (\ref{eq:p^r_i,j(tau)})}\\
q &=& p^{(r)}_{B,G}(T_r) \textrm{\qquad - given by (\ref{eq:p^r_i,j(tau)})}
\end{eqnarray*}

\subsection{The effective loss rate $\pi_B^*$ for non-systematic multipath FEC}\label{appendix:non-systematic}
All formulas shown so far were derived for the systematic version of FEC. The non-systematic FEC$(n,k)$ is easier to handle, and leads to a simplification of these formulas, as follows.

For an \emph{arbitrary schedule} the derivation of (\ref{eq:pi_B^* brute force}) is the same, except that now the number $D(c)$ of lost data packets for a given failure configuration $c$ is
$$D(c) = \left\{ \begin{array}{ll}
    0 & \textrm{ if $\sum_{i=1}^n 1_{\{c_i=B\}}\leq n-k$}\\
    k& \textrm{ otherwise}
    \end{array} \right. $$

Consider now the \emph{equal spacing} on paths. As the number of lost data packets is always $k$ for at least $n\!-\!k\!+\!1$ lost FEC packets, the formula (\ref{eq:BIG pi_B*}) for the effective loss rate $\pi_B^*$ gets simplified to
\begin{eqnarray*}
&&\hspace{-0.5cm} \pi_B^*\ =\ \frac{1}{k} \sum_{j=n-k+1}^n k\cdot \pr(F=j)=\\
&& = \sum_{j=n-k+1}^n  \!\!\sum_{\scriptsize\begin{tabular}{c} $0\leq j_1,\dots,j_R\leq j$ \\$j_1+\ldots+j_R=j$\end{tabular}}\hspace{-1cm} \pr(F_1\!=\!j_1,\ldots,F_R\!=\!j_R) = \\
&& = \sum_{j=n-k+1}^n  \sum_{\scriptsize\begin{tabular}{c} $0\leq j_1,\dots,j_R\leq j$ \\$j_1+\ldots+j_R=j$\end{tabular}} \prod_{r=1}^R \pr(F_r=j_r) = \\
&& =\ \sum_{j=n-k+1}^n  \sum_{\scriptsize\begin{tabular}{c} $0\leq j_1,\dots,j_R\leq j$ \\$j_1+\ldots+j_R=j$\end{tabular}} \prod_{r=1}^R\ldots \\
&&\ \ \ldots \Big(\pi_G^{(r)}\cdot \pr([{\textstyle { n_r-1 \atop j_r}}]\ |\ G) +\ \pi_B^{(r)}\cdot \pr([{\textstyle {n_r-1 \atop j_r-1}}]\ |\ B)\Big),
\end{eqnarray*}
where $\pr([{\textstyle {a \atop b}}]\ |G)$ or $\pr([{\textstyle {a \atop b}}]\ |B)$ are calculated in~\ref{appendix:Recursive equations}.

\bibliographystyle{IEEE}

\end{document}